\documentclass[final,5p]{elsarticle}
\usepackage{lineno,hyperref}

\journal{Acta Materialia}
\usepackage{gensymb}
\usepackage{amsfonts}
\usepackage{siunitx}
\usepackage{amsmath}
\usepackage{amssymb}
\usepackage{graphicx}
\usepackage{dcolumn}
\usepackage{euscript}
\usepackage{multirow}
\usepackage{color}
\usepackage{dblfloatfix}

\graphicspath{{./figs/}}

\providecommand{\U}[1]{\protect\rule{.1in}{.1in}}
\providecommand{\U}[1]{\protect\rule{.1in}{.1in}}

\newcommand{\wlfig}{0.42\textwidth}
\newcommand{\wfig}{0.44\textwidth}
\newcommand{\Wfig}{0.70\textwidth}

\newcommand{\ala}{\alpha_1}
\newcommand{\alb}{\alpha_2}

\newcommand{\ms}{M_\text{s}}

\newcommand{\hcj}{H_\text{cj}}
\newcommand{\bhmax}{BH_\text{max}}
\newcommand{\br}{B_\text{r}}

\newcommand{\nFA}{Non-MA}
\newcommand{\smpl}[1]{\mathbb{#1}}

\newcommand{\rFig}[1]{Figure~\ref{#1}}
\newcommand{\rfig}[1]{Fig.~\ref{#1}}
\newcommand{\rtbl}[1]{Table~\ref{#1}}

\biboptions{sort&compress} 

\begin{document}
\begin{frontmatter}
  
\title{Microstructural and magnetic property evolution with different
  heat-treatment conditions in an alnico alloy}

\author[ames]{Lin Zhou\corref{mycorrespondingauthor}}
\cortext[mycorrespondingauthor]{Corresponding author}
\ead{linzhou@ameslab.gov}
\author[ames]{Wei Tang}
\author[ames]{Liqin Ke}
\author[oak]{Wei Guo}
\author[oak]{Jonathan D. Poplawsky}
\author[ames]{Iver E. Anderson}
\author[ames]{Matthew J. Kramer}
\address[ames]{Ames Laboratory, U.S. Department of Energy, Ames, Iowa 50011, USA}
\address[oak]{Center for Nanophase Materials Sciences, Oak Ridge National Laboratory, Oak Ridge, TN 37831, USA}

\begin{abstract}
Further property enhancement of alnico, an attractive near-term,
non-rare-earth permanent magnet alloy system, primarily composed of
Al, Ni, Co, and Fe, relies on improved morphology control and size
refinement of its complex spinodally decomposed nanostructure that
forms during heat-treatment. Using a combination of transmission
electron microscopy and atom probe tomography techniques, this study
evaluates the magnetic properties and microstructures of an isotropic
32.4Fe-38.1Co-12.9Ni-7.3Al-6.4Ti-3.0Cu (wt.\si{\percent}) alloy in
terms of processing parameters such as annealing temperature,
annealing time, application of an external magnetic field, as well as
low-temperature ``draw'' annealing. Optimal spinodal morphology and
spacing is formed within a narrow temperature and time range
($\sim$\SI{840}{\celsius} and \SI{10}{\minute}) during
thermal-magnetic annealing (MA). The ideal morphology is a mosaic
structure consisting of periodically arrayed $\sim$\SI{40}{\nm}
diameter (Fe-Co)-rich rods ($\ala$ phase) embedded in an (Al-Ni)-rich
($\alb$ phase) matrix. A Cu-enriched phase with a size of
$\sim$\SIrange{3}{5}{\nm} is located at the corners of two adjacent
$\{110\}$ facets of the $\ala$ phase. The MA process significantly
increased remanence ($\br$) ($\sim$\SIrange{40}{70}{\percent}) of the
alloy due to biased elongation of the $\ala$ phase along the
$\langle100\rangle$ crystallographic direction, which is closest in
orientation to the applied magnetic field. The optimum magnetic
properties of the alloy with an intrinsic coercivity ($\hcj$) of
\SI{1845}{Oe} and a maximum energy product ($\bhmax$) of
\SI{5.9}{MGOe} were attributed to the uniformity of the mosaic
structure.
\end{abstract}

\end{frontmatter}


\section{Introduction}
Permanent magnets (PM) are widely used in daily life,
e.g. loudspeakers, electric motors in air conditioners, washing
machines, and hybrid cars. PMs are also used in technological
applications, such as traveling wave tubes, Hall-effect sensors, and
sorting or separation
equipment~\cite{mccallum.arms2014,kramer.cms2012}. The PM family is
mainly composed of ferrites, rare-earth based PM (Nd-Fe-B and Sm-Co),
and alnico~\cite{mccallum.arms2014,kramer.cms2012}. Alnico is an alloy
system composed mainly of Al, Ni, Co, Fe and a small amount of Ti and
Cu. Alnico magnets have attracted recent attention due to their good
performance at high temperature and worldwide abundance of their
elemental
components~\cite{mccallum.arms2014,kramer.cms2012,anderson.jap2015}. Alnico
was discovered in the 1930s and had been continuously developed and
optimized until the 1970s. Currently, the commercial alnico alloys
with the highest coercivity and energy product ($\bhmax$) are alnico
8H with coercivity ($\hcj$) of \SI{2.02}{kOe} and alnico 9 with a
$\bhmax$ of \SI{10.0}{MGOe},
respectively~\cite{kramer.cms2012,zhou.acta2014}. The commercial
alloys consist of periodically distributed and crystallographically
oriented (Fe-Co)-rich ($\ala$) rods embedded in an (Al-Ni)-rich
($\alb$) matrix form by spinodal decomposition (SD) within an applied
magnetic field~\cite{zhou.acta2014}. Microstructural details depend on
specific alloy chemistry and processing~\cite{zhou.acta2014}. Shape
anisotropy of the $\ala$ rods is currently understood to provide a
basic level of coercivity to the alloys~\cite{aquino.msf1999}. The
relatively high magnetic energy product of alnico 9 is derived from a
directional solidification process used to cast the magnet that
provides overall microstructural alignment and magnetic hysteresis
loop ``squareness''.  Theoretical simulations performed for similar
systems suggest that alnico may be able to double its energy product
if the $\ala$ rods can be controlled with a volume fraction of $2/3$
and a diameter of
$\sim$\SIrange{5}{15}{\nm}~\cite{zeng.prb2002,skomski.jap2010}.

Alnico alloys achieve their best magnetic properties after a
complicated heat-treatment (HT) process, which includes a high
temperature ($\sim$\SI{1250}{\celsius}) solution treatment, followed
by short-time thermal-magnetic annealing ($\sim$\SI{800}{\celsius})
and a long-time lower temperature stepped drawing
($\sim$\SIrange{550}{700}{\celsius})~\cite{mccurrie.hfm1982}. Previous
studies indicated that the magnetic properties of alnico are sensitive
to the time and temperature of the magnetic field annealing (MA), as
well as to the constituent elements of the alloy, including minor
additions
~\cite{zhou.acta2014,mccurrie.hfm1982,szymura.actapol1975,sergeyev.ieeetm1970,sergeyev.jap1969,iwama.tjim1974,iwama.tjim1970,iwama.tjim1967,tang.ieeetm2015}.
However, a comprehensive investigation on how different HT parameters
affect the microstructure and magnetic property evolution of alnico is
lacking. Better understanding of the effect of the various processing
steps on morphology and chemistry of the SD phase separation process
for alnico is required in order to achieve the alloy’s theoretical
limit. In this study, we have delved further into unraveling the
complex nanostructures that control the alloy properties by careful
microstructural evaluation of an isotropic alnico alloy with different
HT process conditions. In particular, we analyzed the effects of
annealing time, annealing temperature, as well as the presence of an
external magnetic field. A combination of transmission electron
microscopy (TEM) and atom probe tomography (APT) techniques were used
to characterize both the morphology and chemistry of phases in the
alloy.

\section{Experimental details}
Using commercial alnico 8H as a model composition, a powder having a
composition of 7.3Al-13.0Ni-38.0Co-32.3Fe-3.0Cu-6.4Ti
(wt.\si{\percent}) was produced by close-coupled gas-atomization. The
spherical gas atomized powders provide a uniform starting
material. Sieved powders, $<$\SI{20}{\um} in diameter, were loaded
into a stainless steel can for hot isostatic pressing (HIP). Before
sealing, the powders were outgassed at \SI{425}{\celsius} for
\SI{2}{\hour} under vacuum (\SI{1.3d-3}{\pascal}). The HIP can was
sealed under vacuum by electron beam welding. The HIP process was
performed under a pressure of \SI{60}{\mega\pascal} at
\SI{1250}{\celsius} for \SI{4}{\hour}.  Center sections of the HIP
consolidated alnico were cut into cylinders (\SI{3}{\mm} diameter and
\SI{8}{\mm} length) using electrical discharge machining. The
resulting alloy was polycrystalline with random grain orientations.

Since the HIP samples undergo slow cooling, the nanostructures were
reset by reheating the cylindrical samples at \SI{1250}{\celsius} for
\SI{30}{\minute} in vacuum (\SI{1.3d-4}{\pascal}) and quenched using
an oil bath. The samples were then annealed between \SI{800}{\celsius}
and \SI{860}{\celsius} for different periods of time (0.5 to
\SI{60}{\minute}) with and without an external applied magnetic field
of \SI{1.0}{\tesla}. Some samples also went through an additional
multi-stepped low temperature annealing or ‘drawing process’ that is
typically used in industry. The first step was at \SI{650}{\celsius}
for \SI{5}{\hour} and the second step was at \SI{580}{\celsius} for
\SI{15}{\hour}. The HT process that includes both \SI{10}{\minute}
magnetic field annealing and multi-stepped drawing is termed ``full
heat treatment (FHT) condition''. To capture the effect of applying an
external field on the morphology of SD phases at the early stage of
phase separation, APT analysis was performed on two samples, which
were annealed at \SI{840}{\celsius} for only \SI{90}{\second} with and
without an external applied field of \SI{1}{\tesla}, followed by a
water quench to room temperature. All other samples were furnace
cooled to room temperature after their HT process. Samples labeled as
$\smpl{A}$ to $\smpl{H}$ were selected for TEM analysis and their
heat-treatment conditions are summarized in \rtbl{tbl:ht-condition}.

\begin{table}[hbtp]
  \footnotesize
  \centering
  \caption{Heat-treatment conditions of samples analyzed by TEM.  Draw
    was performed for \SI{5}{\hour}@\SI{650}{\celsius} followed by
    \SI{15}{\hour}@\SI{580}{\celsius}.}
\label{tbl:ht-condition}
\bgroup
\def\arraystretch{1.2}
\begin{tabular}[0.9\linewidth]{ccccccc}
  \hline
  \\[-1.1em]
\multirow{3}{*}{Samples}&  & \multicolumn{3}{c}{MA}                            & & \multirow{3}{*}{Draw} \\ \cline{3-5}
                        &  & Temperature            & Time           & Magnetic & & \\
                        &  & (\si{\celsius})        & (\si{\minute}) & Field    & &        \\
\\[-1.1em]
\hline
\\[-1.2em]
$\smpl{A}$ & & 800 & 10  & Yes & &  \\
$\smpl{B}$ & & 840 & 10  & Yes & &  \\
$\smpl{C}$ & & 860 & 10  & Yes & &  \\
\\[-1.2em]
$\smpl{D}$ & & 800 & 10  & Yes & & Yes \\
$\smpl{E}$ & & 840 & 10  & Yes & & Yes \\
$\smpl{F}$ & & 860 & 10  & Yes & & Yes \\
\\[-1.2em]
$\smpl{G}$ & & 840 & 1.5 & Yes & &  \\
$\smpl{H}$ & & 840 & 60  & Yes & &  \\
\\[-1.2em]
$\smpl{I}$ & & 800 & 10  &     & &  \\
$\smpl{J}$ & & 840 & 10  &     & &  \\
$\smpl{K}$ & & 860 & 10  &     & &  \\
\\[-1.2em]
\hline
\end{tabular}
\egroup
\end{table}

The magnetic properties were measured using a Laboratorio
Elettrofisico Engineering Walker LDJ Scientific AMH-5 Hysteresis graph
with a \SI{5}{\mm} coil and a maximum applied field of \SI{15.0}{kOe}
at room temperature in a closed-loop setup. TEM samples were prepared
by mechanical wedge-polishing followed by a short time, low-voltage Ar
ion-milling with a liquid-nitrogen-cold stage. TEM analysis was
performed on transverse (observation along the magnetic field
direction during annealing) and longitudinal (observation
perpendicular to the magnetic field direction during annealing)
orientations. An FEI Tecnai F20 (\SI{200}{\kV}, equipped with a
field-emission gun (FEG)) was used for microstructural
characterization. Scanning electron microscopy (SEM) was performed
using an FEI Nova FEG-SEM.  For the MA sample, APT tips were lifted
out from a grain with its $[001]$ crystallographic direction parallel
to the external magnetic field direction. The orientation of that
grain was identified using orientation imaging microscopy (OIM) on an
Amray 1845 FEG-SEM. An FEI Nova 200 dual-beam focused ion beam (FIB)
instrument was used to perform lift-outs and annular milling of
targeted grains to fabricate the needle-shaped APT specimens. A wedge
shaped lift-out geometry was used to mount multiple samples on a Si
``microtip'' array to enable the fabrication of several needles from
one wedge lift-out~\cite{thompson.um2007}. APT was performed with a
local electrode atom probe (CAMECA LEAP 4000X HR). Samples were run in
voltage mode with a base temperature of \SI{40}{\K} and
\SI{20}{\percent} pulse fraction at a repetition rate of
\SI{200}{\kHz}. The datasets were reconstructed and analyzed using the
IVAS 3.6.12 software (CAMECA
Instruments)~\cite{miller.sia2007,guo.nanotech2016}.

\section{Experimental results}
\subsection{Effect of magnetic field annealing temperature}

\begin{figure}[!bh]
\centering \includegraphics[width=\wlfig,clip]{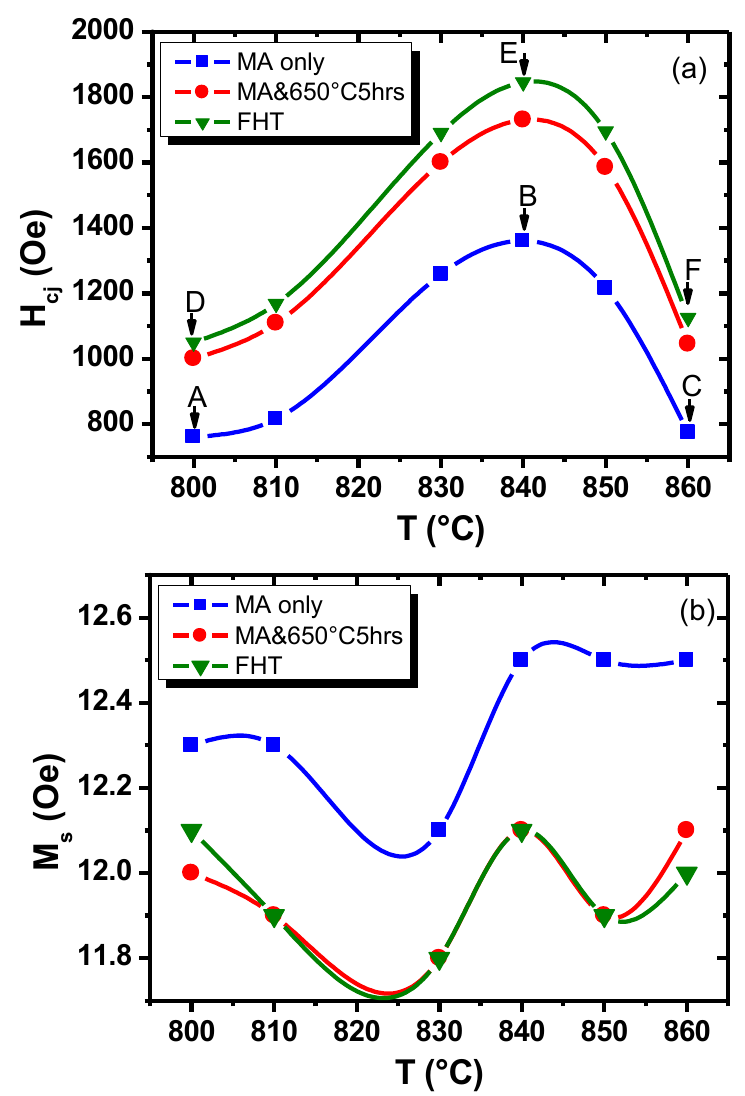}%
\caption{ Effect of MA temperature on (a) $\hcj$ and (b) $\ms$ of
  MA-only (blue), MA+draw$@$\SI{650}{\celsius} (red), and FHT (green)
  samples.}
\label{fig:01}
\end{figure}

\begin{figure*}[!thb]
\centering
\includegraphics[width=\Wfig,clip]{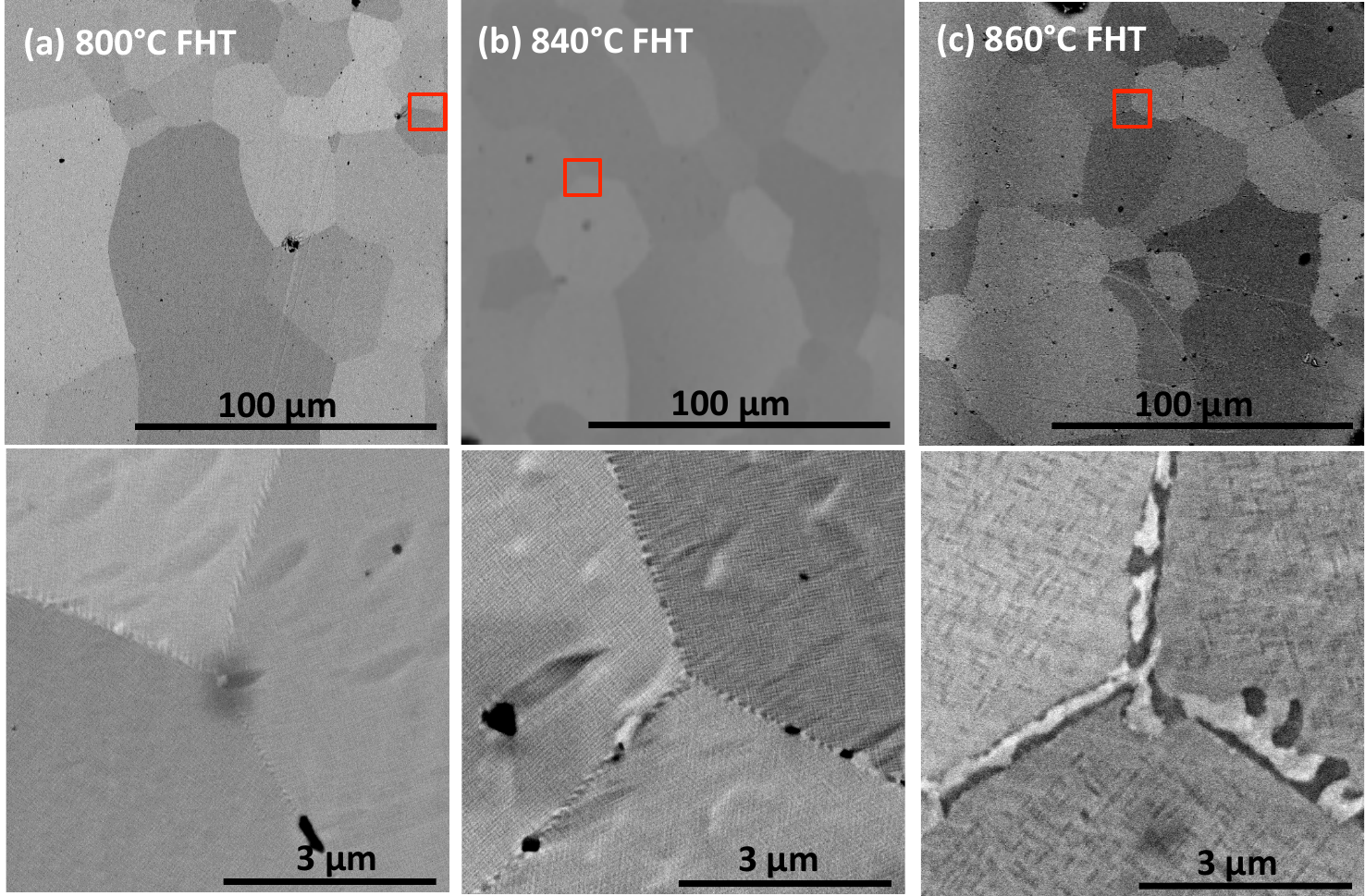}%
\caption{ SEM images of samples (a) $\smpl{D}$, (b) $\smpl{E}$, and
  (c) $\smpl{F}$. The bottom row is a corresponding higher
  magnification image from the image on top.}
\label{fig:02}
\end{figure*}

\begin{figure*}[!bht]
\centering
\includegraphics[width=\Wfig,clip]{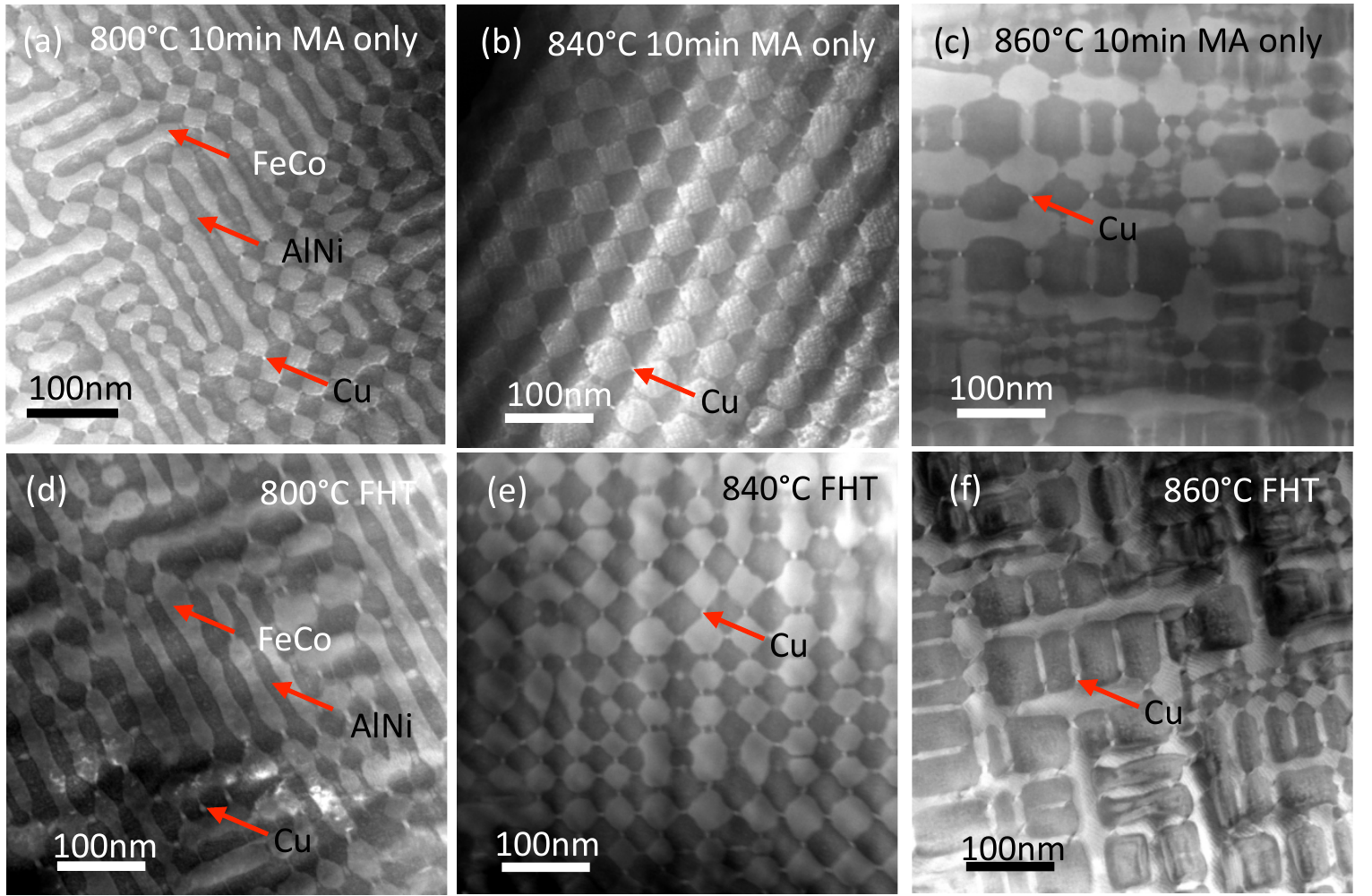}%
\caption{HAADF STEM images of samples (a) $\smpl{A}$, (b) $\smpl{B}$,
  (c) $\smpl{C}$, (d) $\smpl{D}$, (e) $\smpl{E}$, and (f) $\smpl{F}$
  looking along the transverse direction. The TEM images are taken
  close to $[100]$ zone axis. A well defined mosaic pattern is only
  shown in sample $\smpl{B}$ and $\smpl{E}$.}
\label{fig:03}
\end{figure*}

The effect of MA temperature on the intrinsic coercivity ($\hcj$) and
saturation magnetization ($\ms$) is shown in \rfig{fig:01}. The $\hcj$
and $\ms$ of MA-only, MA + \SI{650}{\celsius} (\SI{5}{\hour} draw),
and FHT samples are indicated by the blue, red, and green lines,
respectively. For all samples, it is obvious that $\hcj$ is very
sensitive to the MA temperature. The $\hcj$ maximizes at \SI{1360}{Oe}
at \SI{840}{\celsius} for MA-only and at \SI{1845}{Oe} after FHT. Note
that low-temperature drawing consistently increases $\hcj$ and the
peak position remains at \SI{840}{\celsius}.  Each of the two
annealing steps results with a nearly constant increase. The first
draw after MA caused a slight decrease of $\ms$, but there is no
obvious further change of $\ms$ after the final drawing. To better
understand how the applied magnetic field and various annealing steps
affect magnetic properties, microstructure evolution of select
samples, labeled as $\smpl{A}$ through $\smpl{F}$ in \rfig{fig:01}
were examined in detail.

SEM images of the FHT samples subjected to MA at
\SIlist{800;840;860}{\celsius} (samples $\smpl{D}$, $\smpl{E}$, and
$\smpl{F}$, respectively), are shown in \rfig{fig:02}. For all
samples, an average grain size of $\sim$\SIrange{30}{100}{\um} was
observed. The variation of contrast between grains is due to their
random orientation. At higher magnification, a small phase fraction of
mixed phases at the grain boundaries (GBs) can be observed. The
thickness and, therefore, volume fraction of these GB phases increase
with increasing MA temperature. Energy dispersive X-ray analysis (EDX)
of sample $\smpl{F}$ shows that the bright region in the grain
boundary is (Fe-Co)-rich, while the gray region is (Al-Ni)-rich.

\begin{figure*}[htb]
\centering
\includegraphics[width=\Wfig,clip]{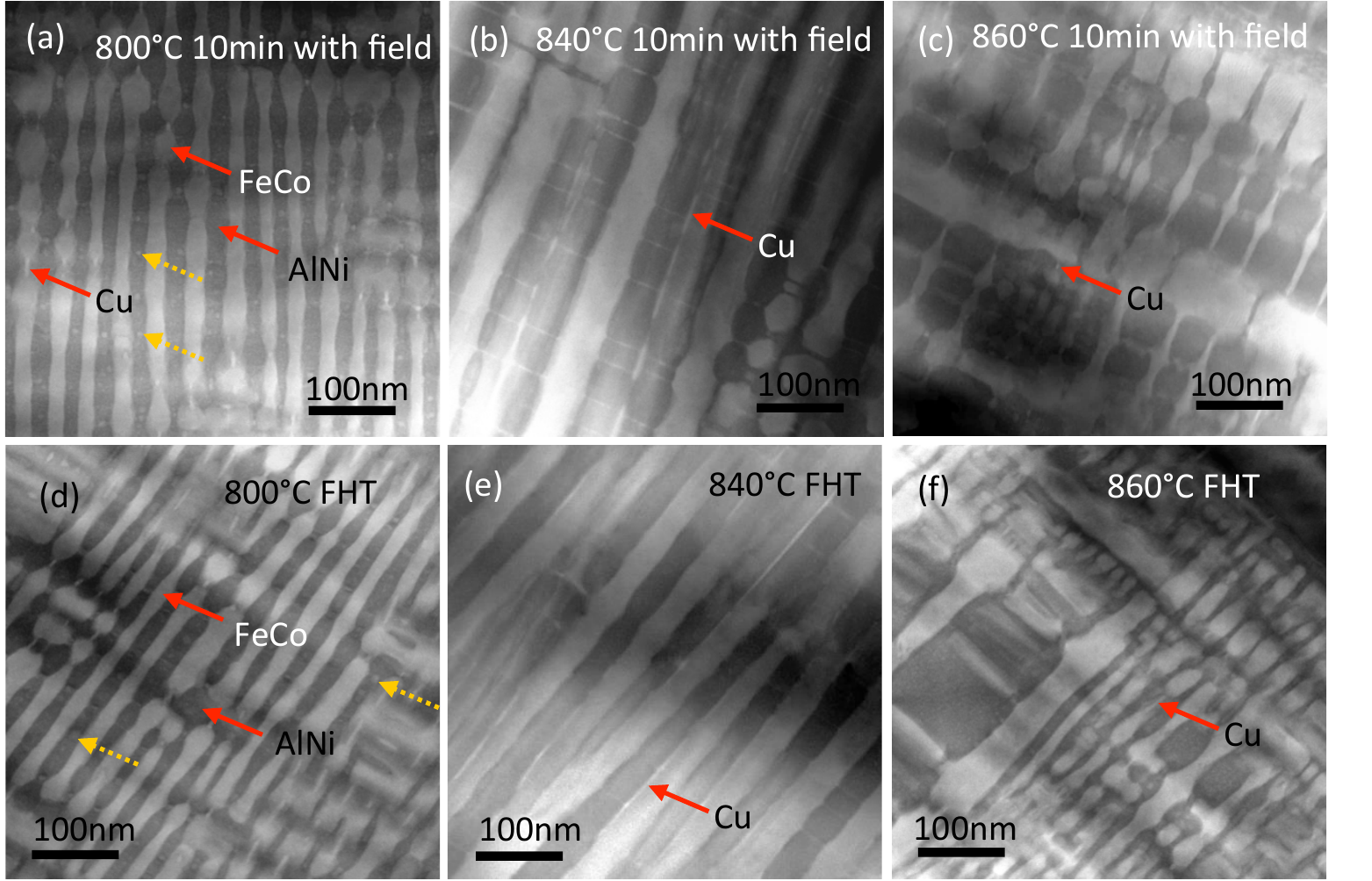}%
\caption{HAADF STEM images of samples (a) $\smpl{A}$, (b) $\smpl{B}$,
  (c) $\smpl{C}$, (d) $\smpl{D}$, (e) $\smpl{E}$, and (f) $\smpl{F}$
  looking along the longitudinal direction. Elongation of the $\ala$
  phase is observed in all samples. Small $\ala$ phases are indicated
  by yellow dotted arrows.}
\label{fig:04}
\end{figure*}

TEM examination of samples $\smpl{A}$ through $\smpl{F}$ was performed
along both the transverse and longitudinal
directions. High-angle-annular-dark-field (HAADF) scanning
transmission electron microscopy (STEM) imaging ($Z$-contrast imaging)
was used to minimize strain contrast and to differentiate phase
morphology. HAADF-STEM imaging is sensitive to the atomic number $Z$,
i.e., elements with higher atomic number will exhibit brighter
contrast in the image because of greater electron scattering at large
detector collection angles~\cite{williams.book1996}. Samples
$\smpl{A}$, $\smpl{B}$, and $\smpl{C}$ were subjected to MA for
\SI{10}{\minute} at \SIlist{800;840;860}{\celsius}, respectively,
followed by furnace cooling. STEM observations along the transverse
direction showed that diamond shaped ($\sim$\SI{20}{\nm} diameter) and
rod shaped ( $\sim$\SI{20}{\nm} diameter and
$\sim$\SIrange{60}{200}{\nm} length) $\ala$ phases coexisted in sample
$\smpl{A}$ (\rfig{fig:03}a). Also, Cu-enriched particles are observed
at the corners of the two $\{110\}$ facets of the diamond shape $\ala$
phases~\cite{zhou.acta2014}. The longer axis of the rod shape $\ala$
phase is aligned along the $\langle100\rangle$ crystallographic
direction but formed nearly orthogonal to the applied magnetic
field. Sample $\smpl{B}$ has a well-defined ``mosaic'' patterned
structure (\rfig{fig:03}b), similar to commercial alnico
8~\cite{zhou.acta2014}. The $\ala$ phase has a diameter of
$\sim$\SIrange{30}{40}{\nm}, indicating an increase in the $\ala$ rod
diameter with a higher annealing temperature. A further increase of MA
temperature to \SI{860}{\celsius} destroys the well-defined mosaic
pattern, as shown in sample $\smpl{C}$ (\rfig{fig:03}c), and the
morphology of $\ala$ phase becomes irregular. Both large $\ala$
regions with sizes up to \SI{70}{\nm} and small $\ala$ phases with a
diameter of $\sim$\SI{10}{\nm} were observed. Cu-enriched particles
were sometimes observed between two $\ala$ phases. Samples $\smpl{D}$
through $\smpl{F}$ were subjected to an additional lower temperature
drawing process (\SI{5}{\hour}@\SI{650}{\celsius} followed by
\SI{15}{\hour}@\SI{580}{\celsius}), after MA at different
temperatures. Compared to the MA-only samples treated at the same
temperature, no significant morphology modification was observed along
the transverse direction after drawing, as shown in
\rfig{fig:03}d--f. However, there are some subtle contrast changes in
the $\ala$ and $\alb$ phases and morphology changes in the Cu-rich
phases as discussed below.

Observation along the longitudinal direction (\rfig{fig:04}) showed
that the $\ala$ phase is elongated for all samples ($\smpl{A}$ through
$\smpl{F}$), though less consistently for the samples subject to MA at
\SI{860}{\celsius}. For sample $\smpl{A}$, the diameter of the
elongated $\ala$ phase is $\sim$\SI{20}{\nm}. Small $\ala$
precipitates ($\sim$\SIrange{3}{5}{\nm}), as indicated by dotted
yellow arrows in \rfig{fig:04}a and confirmed by EDS point analysis,
were commonly observed inside the $\alb$ phase between two elongated
$\ala$ rods. Small $\ala$ precipitates agglomerate and form slightly
bigger particles after the drawing process, as exhibited by sample
$\smpl{D}$ (\rfig{fig:04}d). For samples treated at
\SI{840}{\celsius}, the $\ala$ rods are $\sim$\SIrange{30}{40}{\nm}
wide and up to \SI{1}{\um} long. Cu-enriched rods with a length of
$\sim$\SIrange{10}{50}{\nm} precipitated along the edge of the $\ala$
rods in sample $\smpl{B}$ (\rfig{fig:04}b). After the lower
temperature drawing process, the length of the Cu-enriched rods
increases to $\sim$\SIrange{200}{300}{\nm}, as shown in sample
$\smpl{E}$ (\rfig{fig:04}e). For the sample subject to MA at
\SI{860}{\celsius} (sample $\smpl{F}$), small $\ala$ particles tended
to coarsen after drawing. The length of $\ala$ rods had a wide size
range between tens of nanometers and up to \SI{1}{\um}, as shown in
\rfig{fig:04}f.

\subsection{Effect of magnetic field annealing time}

\begin{figure}[h]
\centering
\includegraphics[width=\wlfig,clip]{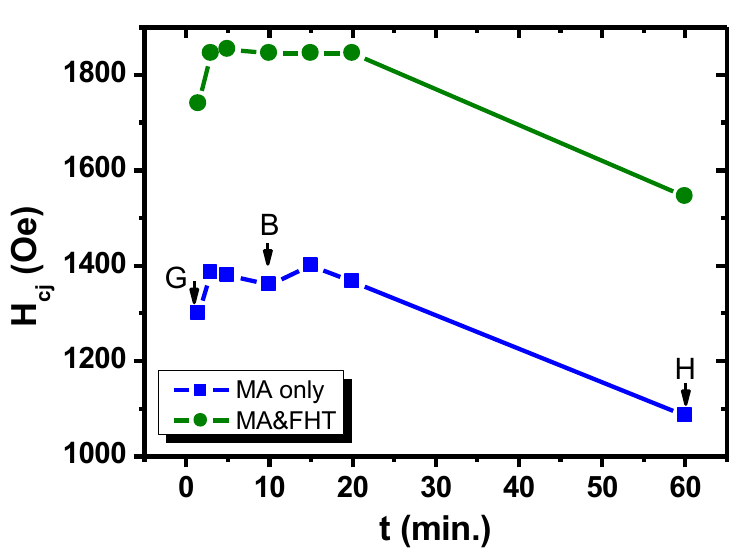}%
\caption{Effect of MA time at \SI{840}{\celsius} on $\hcj$ of MA-only
  (blue) and FHT (green) samples.}
\label{fig:05}
\end{figure}

\begin{figure*}[h]
\centering
\includegraphics[width=\Wfig,clip]{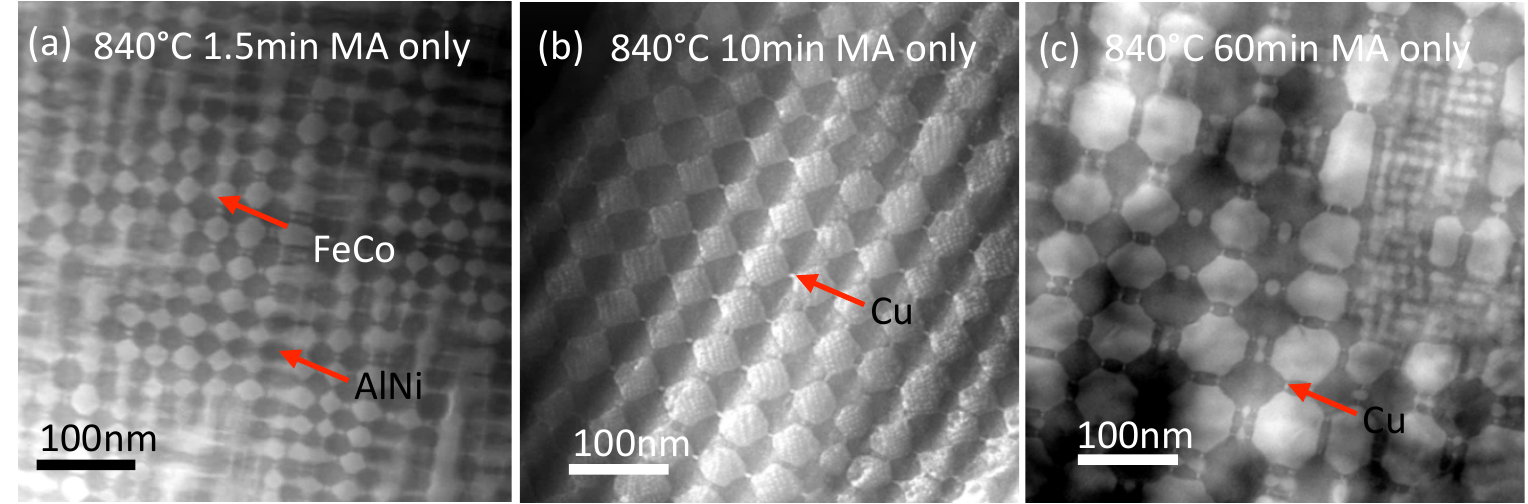}%
\caption{HAADF STEM images of samples (a) $\smpl{G}$, (b) $\smpl{B}$,
  and (c) $\smpl{H}$ looking along the transverse direction. The TEM
  images are taken close to the [100] zone axis. }
\label{fig:06}
\end{figure*}

Since MA at \SI{840}{\celsius} is critical for the formation of a
well-defined mosaic pattern that produces maximum $\hcj$, experiments
were designed to study the effect of the MA time at \SI{840}{\celsius}
on the magnetic properties. Samples were treated at \SI{840}{\celsius}
for time ranging from \SI{30}{\second} to \SI{60}{\minute}, followed
by furnace cooling. The measured $\hcj$ values are shown in
\rfig{fig:05}. The blue line denotes samples with the MA treatment
only, while the green line denotes samples treated with the FHT
process. The $\hcj$ first increases to $\sim$\SI{1400}{Oe} as MA time
increases from $0$ to $\sim$\SI{3}{\minute}. Between $3$ and
\SI{20}{\minute}, the $\hcj$ stays relatively constant. After a
\SI{60}{\minute} MA, the $\hcj$ decreases to
$\sim$\SI{1100}{Oe}. Drawing improves the $\hcj$ of all samples and
the increase of $\hcj$ with low temperature annealing for the drawn
samples follows the same trend as that for samples without
drawing. Samples labeled as $\smpl{G}$ and $\smpl{H}$ in \rfig{fig:05}
were selected for further TEM analysis.

\rFig{fig:06} shows the SD morphologies viewed along the transverse
direction of samples $\smpl{G}$, $\smpl{B}$ (same as in Fig.3 but used
again here for comparison), and $\smpl{H}$ with MA time of
\SIlist{1.5;10;60}{\minute}, respectively. The diameter of the $\ala$
phase increases from $\sim$\SI{15}{\nm} to $\sim$\SI{35}{\nm} with
increasing MA time from \SI{1.5}{\minute} to \SI{10}{\minute},
respectively. Most of the area imaged in sample $\smpl{G}$
(\rfig{fig:06}a) formed a mosaic pattern, but some regions exhibited a
blurry contrast, which appears to originate from incomplete phase
separation. Increasing the MA time to \SI{60}{\minute} drastically
modifies the SD morphology, as shown in \rfig{fig:06}c (sample
$\smpl{H}$). The size of the $\ala$ phase transformed into a bimodal
distribution during the extended MA. The morphology consists of a
larger $\ala$ phase with a size of $\sim$\SIrange{60}{80}{\nm} and a
smaller round (non-faceted) $\ala$ phase with a size of
$\sim$\SI{10}{\nm}. The larger $\ala$ phase has alternating $\{100\}$
and $\{110\}$ facets.  Some of the smaller $\ala$ phases are located
at the regions between two $\{100\}$ facets of the bigger $\ala$
phases and some of them form clusters. Cu-enriched phases were also
observed at the corners of two larger $\ala$ facets, but these
Cu-enriched phases are slightly smaller compared to those in sample
$\smpl{B}$.

\subsection{Effect of magnetic field}

\begin{figure}[!htb]
\centering
\includegraphics[width=\wlfig,clip]{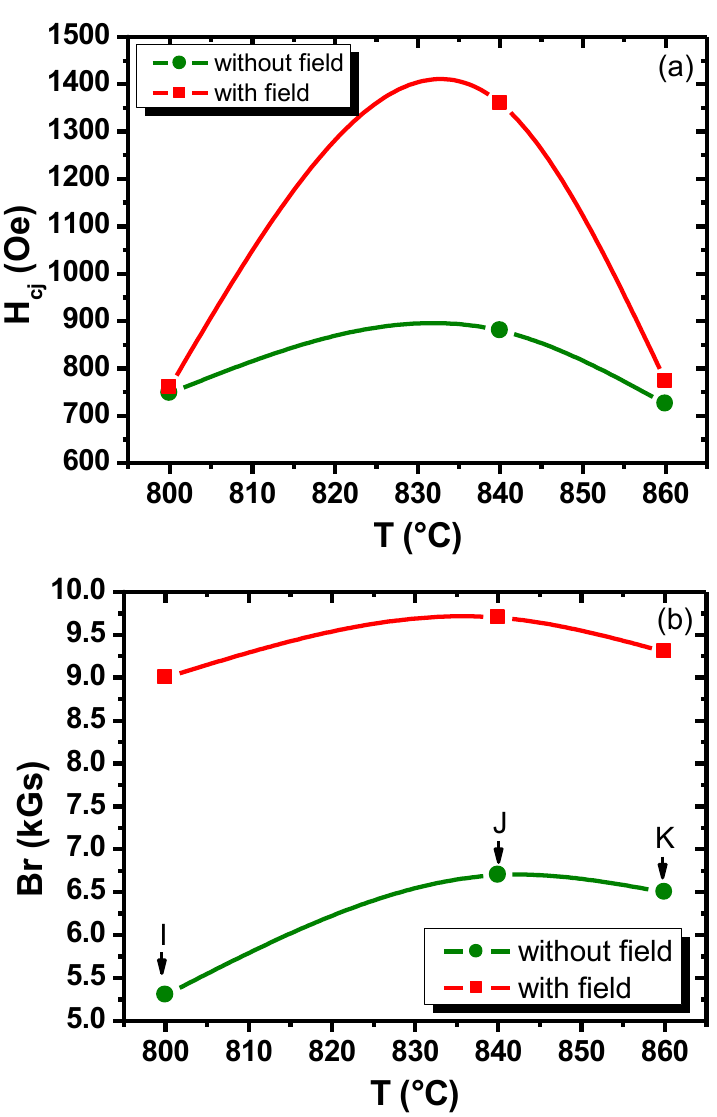}%
\caption{Comparison on effect of magnetic field on (a) $\hcj$ and (b)
  $\br$ of samples treated at different temperatures. Red line are
  from samples treated with a magnetic field, while the green line are
  from samples treated without a magnetic field.}
\label{fig:07}
\end{figure}

Finally, the magnetic field effect was analyzed by comparing the
magnetic properties ($\hcj$ and $\br$) of samples treated with/without
external magnetic fields at different temperatures, as shown in
\rfig{fig:07}. Samples treated in the absence of a magnetic field at
\SIlist{800;840;860}{\celsius} are labeled as $\smpl{I}$, $\smpl{J}$,
and $\smpl{K}$, respectively, and are compared to samples $\smpl{A}$,
$\smpl{B}$, and $\smpl{C}$ in \rfig{fig:07}. For both sets of samples,
there is a maximum $\hcj$ at \SI{840}{\celsius}, but the $\hcj$ of
sample $\smpl{J}$ is $\sim$\SI{840}{Oe}, which is smaller than that of
sample $\smpl{B}$. The presence of an external magnetic field barely
affects the $\hcj$ of samples treated at 800 and
\SI{860}{\celsius}. On the other hand, all samples treated using a
magnetic field have an obviously higher $\br$
($\sim$\SI{45}{\percent}) compared with those treated without a
magnetic field, as shown in \rfig{fig:07}b.

\begin{figure*}[htb]
\centering
\includegraphics[width=\Wfig,clip]{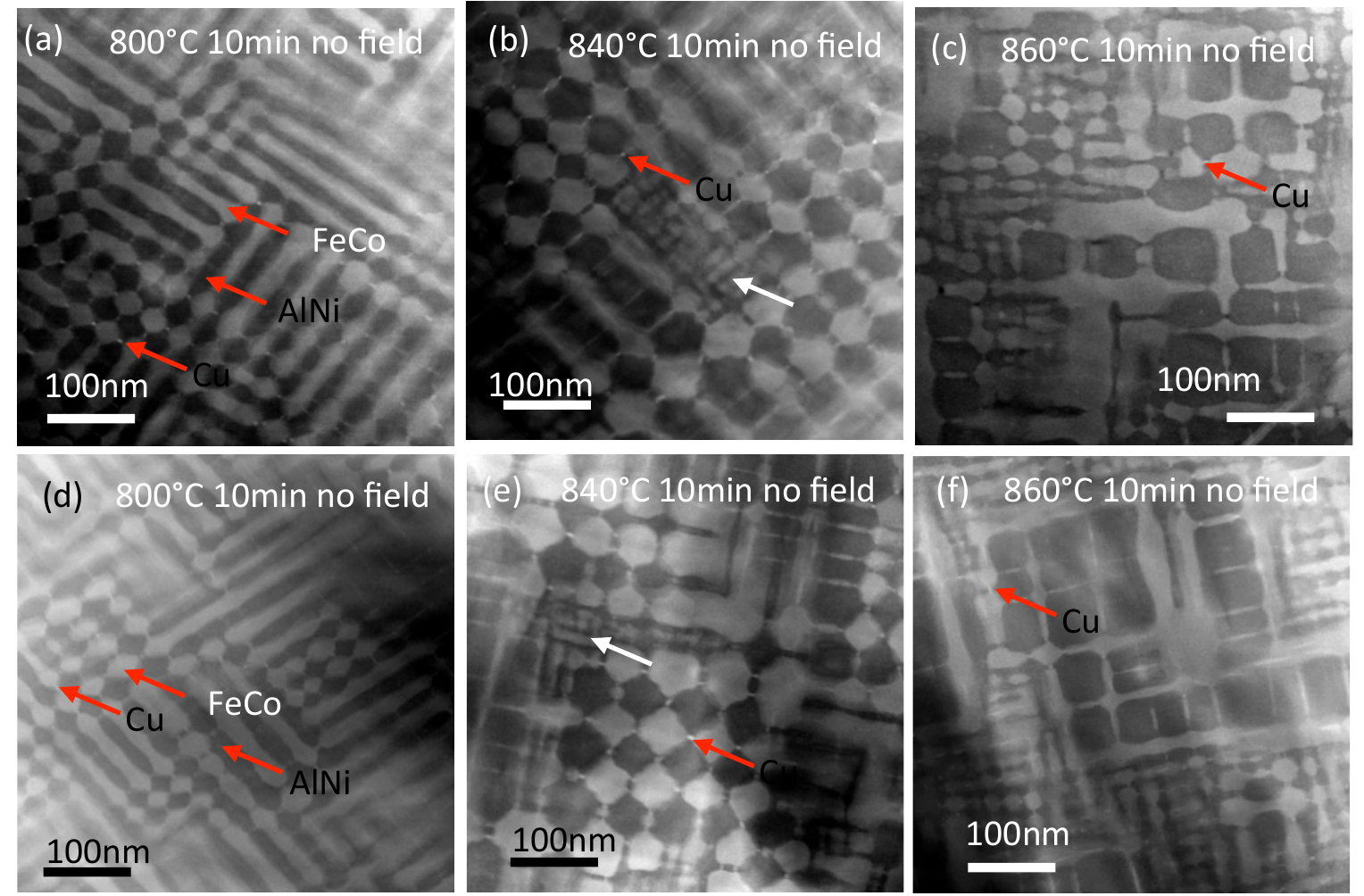}%
\caption{ Transverse HAADF STEM images of samples $\smpl{I}$ (a),
  $\smpl{J}$ (b), $\smpl{K}$ (c), and longitudinal HAADF STEM images
  of samples $\smpl{I}$ (d), $\smpl{J}$ (e), $\smpl{K}$ (f).}
\label{fig:08}
\end{figure*}

TEM analysis along the transverse direction shows that the morphology
of the SD phases in samples $\smpl{I}$ and $\smpl{K}$ (\rfig{fig:08}a
and ~\ref{fig:08}c, respectively) is similar to that of samples
$\smpl{A}$ and $\smpl{C}$ (\rfig{fig:03}a and ~\ref{fig:03}c),
indicating no biasing of the microstructure along the transverse
direction if MA is performed away from the optimum temperature
(\SI{840}{\celsius}). However, a SD morphology difference
(\rfig{fig:03}b and \rfig{fig:08}b) was observed between sample
$\smpl{B}$ and sample $\smpl{J}$. Only some areas of sample $\smpl{J}$
exhibit a mosaic pattern. The faceting of $\ala$ rods is not as well
defined as those in sample $\smpl{B}$. Moreover, between regions with
a mosaic pattern, there are clusters of the $\ala$ phase with smaller
diameters, as indicated by the white arrow in \rfig{fig:08}b. Viewing
samples along the longitudinal direction showed that the SD phase
morphology is similar to that observed along the transverse direction,
which obviously suggests that the $\ala$ phases in samples
($\smpl{I}$, $\smpl{J}$, and $\smpl{K}$) are oriented isotropically,
as expected without an external field biasing effect.

\begin{figure}[bht]
\centering
\includegraphics[width=0.48\textwidth,clip]{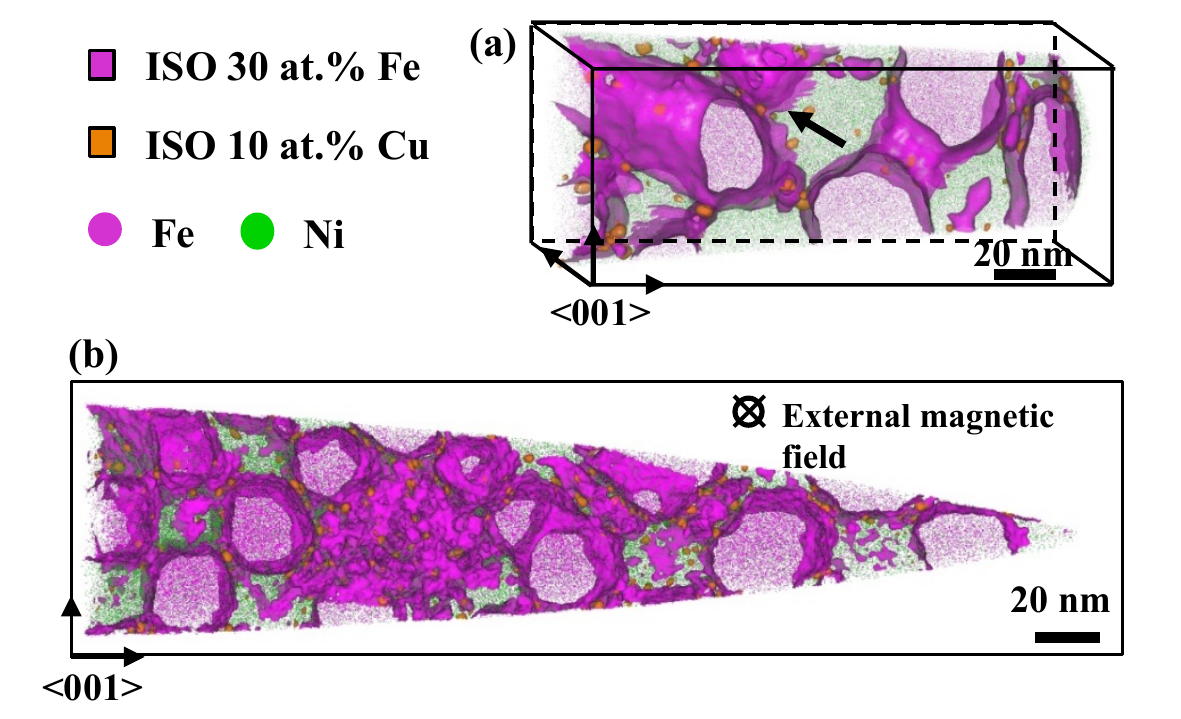}%
\caption{APT reconstruction of the alnico 8 sample annealed at
  \SI{840}{\celsius} for \SI{90}{\second} with (a) an external
  magnetic field and without (b) an external magnetic field. The
  initial nanostructures evolution is viewed in perpendicular
  $\langle001\rangle$ directions, together with 30 at.\si{\percent} Fe
  isoconcentration surfaces and 10 at.\si{\percent} Cu
  iso-surfaces. The dark arrow in (a) indicates two orthogonally
  aligned $\ala$ rods.  }
\label{fig:09}
\end{figure}

The effects of external magnetic field on the morphology of SD phases
at the early stage of SD are shown in the 3D APT reconstructed
iso-surfaces in \rfig{fig:09}. The non-field annealed (\nFA) sample
has $\ala$ rods with their longer axis aligned along all three
orthogonal $\langle100\rangle$ directions
(\rfig{fig:09}a)~\cite{guo.mm2016}. In contrast, in the MA sample, the
$\ala$ phase is much longer and preferentially elongates along the
$\langle100\rangle$ direction closest to the external magnetic field
direction (\rfig{fig:09}b).  The APT bulk composition from the center
of the $\ala$ and $\alb$ phases of both samples
(\rtbl{tbl:composition}) shows similar concentrations in the two
$\ala$ and $\alb$ phases, indicating that the field mainly biases the
morphology of the SD phases and has no obvious effects on their
chemistry.

\begin{table}[hbtp]
  \footnotesize
  \centering
\caption{Phase compositions of the $\ala$ and $\alb$ phases in MA and
  {\nFA} samples.}
\label{tbl:composition}
\bgroup
\def\arraystretch{1.2}
\begin{tabular}[0.9\linewidth]{clcccccc}
  \hline
  \\[-1.1em]
Phase & Sample & Al & Ti & Fe & Co & Ni & Cu\\
\\[-1.1em]
\hline
\\[-1.2em]
\multirow{2}{*}{$\ala$} & MA     & 4.28  & 1.01  & 52.47 & 36.48 & 4.55 & 1.22\\
                        & \nFA   & 4.35  & 1.04  & 51.92 & 36.65 & 4.79 & 1.22\\
\multirow{2}{*}{$\alb$} & MA     & 22.45 & 14.98 & 13.08 & 33.36 & 14.29 & 1.86\\
                        & \nFA   & 22.06 & 14.52 & 13.72 & 33.08 & 14.69 & 1.89\\
\\[-1.2em]
\hline
\end{tabular}
\egroup
\end{table}

\section{Discussion}
The results presented clearly show that the magnetic properties of
alnico are closely related to the morphology of the SD phases, which
in turn are sensitive to both thermal MA and post-MA processing. The
best magnetic properties are promoted by well-aligned rods of the
$\ala$ phase in a uniformly spaced mosaic pattern. This pattern is
observed only if a magnetic field is applied in a very narrow
temperature range for a limited time (e.g. $\sim$\SI{840}{\celsius}
for \SI{10}{\minute} for this alloy). Subsequent low temperature
anneals are required to further increase the coercivity, which can be
increased as much as \SI{30}{\percent}. There are no apparent
morphological changes corresponding to the increase in the coercivity,
except for the elongation of the Cu-enriched phase.

These results suggest that there is likely several mechanisms at play
in addition to SD. For alloys with a miscibility gap, a two-phase
mixture can be formed by either classical nucleation and growth (NG)
or by an SD path~\cite{laughlin.asmhandbook1985}. The NG path arises
with small undercooling (low supersaturation) and requires relatively
large localized composition
fluctuations~\cite{laughlin.asmhandbook1985}. The size of the
nucleated second phase grows with phase separation time. The SD path
occurs at large undercooling and the transformation occurs
homogeneously throughout the alloy by continuous growth of initially
small compositional
fluctuations~\cite{laughlin.asmhandbook1985}. Theoretical studies on
magnetic aging of the SD alloy showed that early in the SD process,
the composition wave length is determined by the interplay of chemical
free energy, elastic energy, and magnetic
energy~\cite{cahn.acta1961,cahn.jap1963}. The elastic energy
suppresses the system SD, while the magnetic energy suppresses the
development of composition waves along the external field
direction~\cite{cahn.jap1963}. The magnetic energy can have a serious
effect on the SD process only when the alloy is annealed just below
the Curie temperature, where the change of magnetic energy is the
largest~\cite{cahn.jap1963}. Our experiments showed that the MA
process is very effective in forming uniformly spaced rods only when
MA is conducted at a specific temperature range
($\sim$\SI{840}{\celsius} for this alloy, samples $\smpl{B}$ and
$\smpl{E}$ in \rtbl{tbl:ht-condition}), which tends to support this
prediction.

\begin{figure}[htb]
\centering
\includegraphics[width=\wfig,clip]{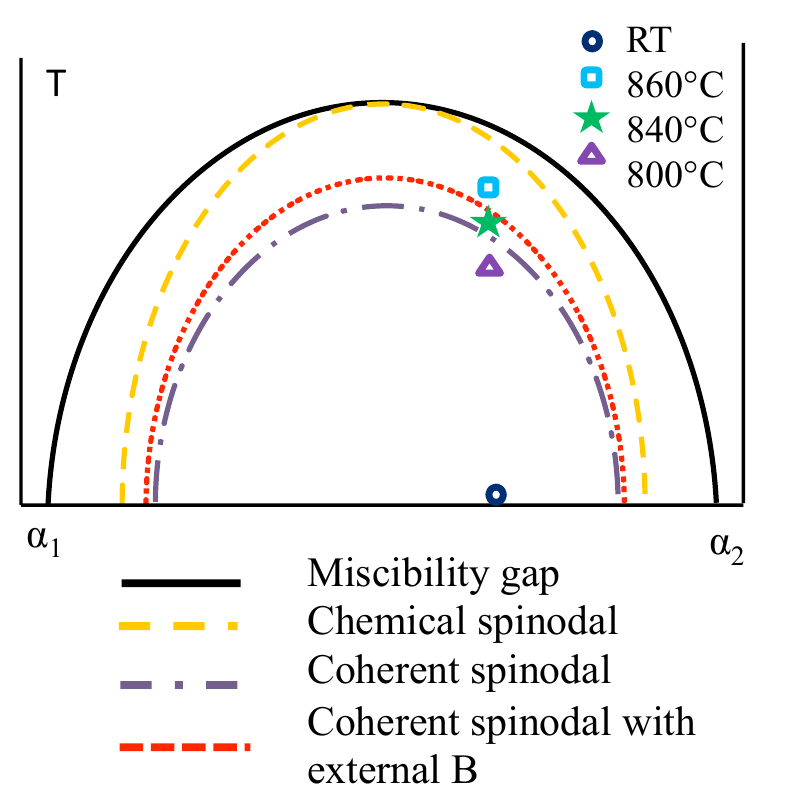}%
\caption{Schematic of phase diagram of alnico 8 by considering it as a
  quasi-binary system with $\ala$ and $\alb$ phases.}
\label{fig:10}
\end{figure}

Considering alnico as a pseudo-binary alloy that consists of $\ala$
and $\alb$ phases with a miscibility gap, its phase diagram may be
illustrated schematically in \rfig{fig:10}. The chemical spinodal line
is defined by the inflexion point of the isothermal free energy ($G$)
composition curve \newline ($\partial G^2/\partial C^2$=$0$).  The
coherent SD line is defined by the displacement of the SD curve due to
strain energy. Since the SD process is homogeneous and the SD phase is
typically uniform within a grain~\cite{laughlin.asmhandbook1985}, it
is reasonable to assume that a homogeneous structure with a periodic
pattern in the alloy implies that the alloy decomposes along the path
of SD. Transverse observations from TEM indicate that samples
$\smpl{A}$, $\smpl{B}$, and $\smpl{I}$ (MA @ \SI{840}{\celsius} or
lower) follow a SD path, while samples $\smpl{C}$, $\smpl{J}$, and
$\smpl{K}$ (MA@ $>$ \SI{840}{\celsius}) follow a NG path. As a result,
the SD coherent spinodal line shifts upward due to application of an
external field, as outlined in \rfig{fig:09}. This may be because the
application of an external magnetic field promotes elongation of the
$\ala$ phase along the field direction substantially at
\SI{840}{\celsius}, which can slightly modify the strain energy of the
system and change the phase separation mechanism perpendicular to the
field direction from NG to SD.

The similarity in the SD morphologies of samples treated with MA only
(samples $\smpl{A}$, $\smpl{B}$, and $\smpl{C}$ in
\rtbl{tbl:ht-condition}) and FHT processes (samples $\smpl{D}$,
$\smpl{E}$, and $\smpl{F}$ in \rtbl{tbl:ht-condition}) reveals that
the shape of the $\ala$ phase is established during the MA stage by
SD. Drawing may further ``tune'' the chemistry differences between the
$\ala$ and $\alb$ phases, which can cause larger magnetization
differences between $\ala$ and $\alb$ phases. This is because at a
much lower temperature (draw annealing), the long-range kinetic
effects during MA are suppressed, but the chemical driving forces
still allow for small, but important chemical separation to
continue. As a result, there is a large increase in $\hcj$ with
FHT. Details on the morphology and chemistry changes, as well as the
evolution of Cu-enriched phase during MA at \SI{840}{\celsius} and the
drawing process, will be reported elsewhere.

The changes of the magnetic properties and microstructures with
different MA time at \SI{840}{\celsius} (samples $\smpl{G}$,
$\smpl{B}$, and $\smpl{H}$) suggest that the mosaic structure set by
the MA process after $\sim$\SI{10}{\minute}, which gives the best
magnetic properties, is a metastable structure. The $\ala$ phase will
grow into a different morphology if the sample is subjected to MA for
a long enough time. The mosaic structure may be temporarily stabilized
by the Cu-enriched phase at the corners of $\ala$ rods, which
introduces a large amount of strain and constrains the growth of the
$\ala$ phase. Observation of the Cu-enriched phase with a smaller size
in sample $\smpl{H}$ tends to support the hypothesis that the $\ala$
phase can grow larger only after the Cu-enriched phase is broken into
smaller particles by extended magnetic annealing.

\begin{figure}[htb]
\centering
\includegraphics[width=\wlfig,clip]{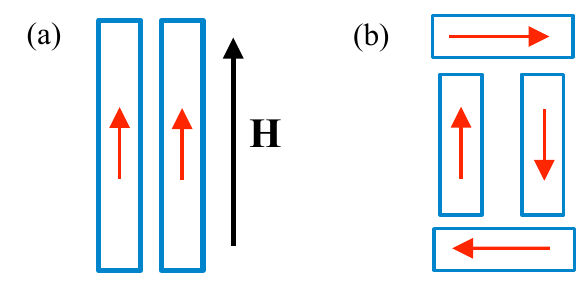}%
\caption{Arrangement of $\ala$ phase with (a) and without (b) an
  external magnetic field. Red arrows indicate the direction of
  magnetic moment inside the $\ala$ phase.}
\label{fig:11}
\end{figure}

Particle elongation during MA seems to be driven by magnetostatic
interaction, which is the same provider of magnetic shape anisotropy
and coercivity in alnico. For shape anisotropy, the shape of the
elongated magnetic particles is fixed and the magnetization tends to
align along the longer axis of the particles to lower magnetostatic
energy. Similarly, for particle elongation during MA, the
magnetization vectors of the magnetic phase are fixed along the field
direction (For simplicity, we assume that a saturated field is applied
along the [001] direction.) and the magnetic phases tend to evolve
into elongated shapes with their long axes along or close to the field
direction.

The two major phases in alnico, $\ala$ and $\alb$, have coherent
lattices and the corresponding interfacial energy is relatively small
~\cite{cahn.jap1963,zijlstra.zap1962,neel.cr1947}. This results in a
high sensitivity of the microstructure to the external magnetic field
during MA, especially at the onset of SD and just below the Curie
temperature, when the composition difference between two phases is
still small, but the magnetization difference may be large. The
composition waves along the field direction are thus effectively
suppressed at the initial stage of SD~\cite{cahn.jap1963}. Thus, a
well-defined mosaic pattern is developed within a short time during MA
(\rfig{fig:11}a) perpendicular to the field direction. After the SD
phases become fully developed, the composition difference between the
two phases and their interfacial energy both increase. For an extended
MA, $\ala$ rods may become larger and more round (with both $\{100\}$
and $\{110\}$ faceting), as shown in \rfig{fig:06}c, to minimize the
interfacial energy. In the absence of an external magnetic field, the
$\ala$ phase may still elongate along a $\langle100\rangle$ direction
and form closed loops to minimize the magnetostatic energy. Elongated
$\ala$ phases embedded in $\alb$ phases still form in three
dimensions, but the overall microstructure is isotropic, as
illustrated in \rfig{fig:11}b.

\section{Conclusions}
The correlations between magnetic properties, microstructure and HT
conditions have been comprehensively studied for alnico. The
morphology of SD phases is very sensitive to MA temperature and time
for the alnico composition in this study. A structure formed by
uniformly spaced, $\sim$\SI{40}{\nm} (diameter), $\{110\}$ faceted
$\ala$ rods in a $\alb$ matrix can only be achieved for a narrow MA
temperature range ($\sim$\SI{840}{\celsius}), and within limited MA
time ($\sim$\SIrange{3}{10}{\minute}), resulting in suitable magnetic
properties, similar to commercial magnets of this type. The external
magnetic field aids the formation of a well-faceted mosaic structure
only at the optimal temperature ($\sim$\SI{840}{\celsius}). MA at
different temperatures causes elongation of the $\ala$ phase as well,
which improves $\br$ of alnico, but has no obvious effect on
$\hcj$. Low temperature drawing can further increase the alloy’s
coercivity by modifying local chemistry in the $\ala$ and $\alb$
phases and their interfaces, without substantially changing the $\ala$
rod diameter and spacing.

\section*{Acknowledgments}
Research was supported by U.S. DOE, Office of Energy Efficiency and
Renewable Energy (EERE), under its Vehicle Technologies Office,
Electric Drive Technology Program, through the Ames Laboratory, Iowa
State University under contract DE-AC02-07CH11358. APT research was
conducted as part of a user project at ORNL’s Center for Nanophase
Materials Sciences (CNMS), which is a DOE Office of Science User
Facility.

\section*{References}

\bibliography{aaa}

\begin{thebibliography}{25}
\expandafter\ifx\csname natexlab\endcsname\relax\def\natexlab#1{#1}\fi
\providecommand{\url}[1]{\texttt{#1}}
\providecommand{\href}[2]{#2}
\providecommand{\path}[1]{#1}
\providecommand{\DOIprefix}{doi:}
\providecommand{\ArXivprefix}{arXiv:}
\providecommand{\URLprefix}{URL: }
\providecommand{\Pubmedprefix}{pmid:}
\providecommand{\doi}[1]{\href{http://dx.doi.org/#1}{\path{#1}}}
\providecommand{\Pubmed}[1]{\href{pmid:#1}{\path{#1}}}
\providecommand{\bibinfo}[2]{#2}
\ifx\xfnm\relax \def\xfnm[#1]{\unskip,\space#1}\fi
\bibitem[{McCallum et~al.(2014)McCallum, Lewis, Skomski, Kramer, and
  Anderson}]{mccallum.arms2014}
\bibinfo{author}{R.~McCallum}, \bibinfo{author}{L.~Lewis},
  \bibinfo{author}{R.~Skomski}, \bibinfo{author}{M.~Kramer},
  \bibinfo{author}{I.~Anderson},
\newblock \bibinfo{title}{Practical aspects of modern and future permanent
  magnets},
\newblock \bibinfo{journal}{Annual Review of Materials Research}
  \bibinfo{volume}{44} (\bibinfo{year}{2014}) \bibinfo{pages}{451--477}.
\bibitem[{Kramer et~al.(2012)Kramer, McCallum, Anderson, and
  Constantinides}]{kramer.cms2012}
\bibinfo{author}{M.~Kramer}, \bibinfo{author}{R.~McCallum},
  \bibinfo{author}{I.~Anderson}, \bibinfo{author}{S.~Constantinides},
\newblock \bibinfo{title}{Prospects for non-rare earth permanent magnets for
  traction motors and generators},
\newblock \bibinfo{journal}{JOM Journal of the Minerals, Metals and Materials
  Society} \bibinfo{volume}{64} (\bibinfo{year}{2012})
  \bibinfo{pages}{752--763}.
\bibitem[{Anderson et~al.(2015)Anderson, Kassen, White, Zhou, Tang, Palasyuk,
  Dennis, McCallum, and Kramer}]{anderson.jap2015}
\bibinfo{author}{I.~E. Anderson}, \bibinfo{author}{A.~G. Kassen},
  \bibinfo{author}{E.~M.~H. White}, \bibinfo{author}{L.~Zhou},
  \bibinfo{author}{W.~Tang}, \bibinfo{author}{A.~Palasyuk},
  \bibinfo{author}{K.~W. Dennis}, \bibinfo{author}{R.~W. McCallum},
  \bibinfo{author}{M.~J. Kramer},
\newblock \bibinfo{title}{Novel pre-alloyed powder processing of modified
  alnico 8: Correlation of microstructure and magnetic properties},
\newblock \bibinfo{journal}{Journal of Applied Physics} \bibinfo{volume}{117}
  (\bibinfo{year}{2015}) \bibinfo{pages}{17D138}.
\bibitem[{Zhou et~al.(2014)Zhou, Miller, Lu, Ke, Skomski, Dillon, Xing,
  Palasyuk, McCartney, Smith, Constantinides, McCallum, Anderson, Antropov, and
  Kramer}]{zhou.acta2014}
\bibinfo{author}{L.~Zhou}, \bibinfo{author}{M.~Miller},
  \bibinfo{author}{P.~Lu}, \bibinfo{author}{L.~Ke},
  \bibinfo{author}{R.~Skomski}, \bibinfo{author}{H.~Dillon},
  \bibinfo{author}{Q.~Xing}, \bibinfo{author}{A.~Palasyuk},
  \bibinfo{author}{M.~McCartney}, \bibinfo{author}{D.~Smith},
  \bibinfo{author}{S.~Constantinides}, \bibinfo{author}{R.~McCallum},
  \bibinfo{author}{I.~Anderson}, \bibinfo{author}{V.~Antropov},
  \bibinfo{author}{M.~Kramer},
\newblock \bibinfo{title}{Architecture and magnetism of alnico},
\newblock \bibinfo{journal}{Acta Materialia} \bibinfo{volume}{74}
  (\bibinfo{year}{2014}) \bibinfo{pages}{224--233}.
\bibitem[{Matutes-Aquino et~al.(1999)Matutes-Aquino,
  Dom{\'{\i}}nguez~R{\'{\i}}os, Miki~Yoshida, and
  Ayala~Valenzuela}]{aquino.msf1999}
\bibinfo{author}{J.~A. Matutes-Aquino},
  \bibinfo{author}{C.~Dom{\'{\i}}nguez~R{\'{\i}}os},
  \bibinfo{author}{M.~Miki~Yoshida}, \bibinfo{author}{O.~Ayala~Valenzuela},
\newblock \bibinfo{title}{Magnetic properties and microstructure of the alnico
  8},
\newblock in: \bibinfo{booktitle}{Magnetism, Magnetic Materials and their
  Applications}, volume \bibinfo{volume}{302} of
  \textit{\bibinfo{series}{Materials Science Forum}}, \bibinfo{publisher}{Trans
  Tech Publications}, \bibinfo{year}{1999}, pp. \bibinfo{pages}{329--333}.
\bibitem[{Zeng et~al.(2002)Zeng, Skomski, Menon, Liu, Bandyopadhyay, and
  Sellmyer}]{zeng.prb2002}
\bibinfo{author}{H.~Zeng}, \bibinfo{author}{R.~Skomski},
  \bibinfo{author}{L.~Menon}, \bibinfo{author}{Y.~Liu},
  \bibinfo{author}{S.~Bandyopadhyay}, \bibinfo{author}{D.~J. Sellmyer},
\newblock \bibinfo{title}{Structure and magnetic properties of ferromagnetic
  nanowires in self-assembled arrays},
\newblock \bibinfo{journal}{Phys. Rev. B} \bibinfo{volume}{65}
  (\bibinfo{year}{2002}) \bibinfo{pages}{134426}.
\bibitem[{Skomski et~al.(2010)Skomski, Liu, Shield, Hadjipanayis, and
  Sellmyer}]{skomski.jap2010}
\bibinfo{author}{R.~Skomski}, \bibinfo{author}{Y.~Liu}, \bibinfo{author}{J.~E.
  Shield}, \bibinfo{author}{G.~C. Hadjipanayis}, \bibinfo{author}{D.~J.
  Sellmyer},
\newblock \bibinfo{title}{Permanent magnetism of dense-packed nanostructures},
\newblock \bibinfo{journal}{Journal of Applied Physics} \bibinfo{volume}{107}
  (\bibinfo{year}{2010}) \bibinfo{pages}{09A739}.
\bibitem[{McCurrie(1982)}]{mccurrie.hfm1982}
\bibinfo{author}{R.~McCurrie},
\newblock \bibinfo{title}{Chapter 3 the structure and properties of alnico
  permanent magnet alloys},
\newblock volume~\bibinfo{volume}{3} of \textit{\bibinfo{series}{Handbook of
  Ferromagnetic Materials}}, \bibinfo{publisher}{Elsevier},
  \bibinfo{year}{1982}, pp. \bibinfo{pages}{107--188}.
\bibitem[{Szymura et~al.(1975)Szymura, Wyslocki, and
  Bailon}]{szymura.actapol1975}
\bibinfo{author}{S.~Szymura}, \bibinfo{author}{B.~Wyslocki},
  \bibinfo{author}{J.~Bailon},
\newblock \bibinfo{title}{Effect of treatment on the domain structures and
  magnetic properties of alnico alloys},
\newblock \bibinfo{journal}{Acta Phys. Polon.}  (\bibinfo{year}{1975})
  \bibinfo{pages}{177--181}.
\bibitem[{Sergeyev and Bulygina(1970)}]{sergeyev.ieeetm1970}
\bibinfo{author}{V.~Sergeyev}, \bibinfo{author}{T.~Bulygina},
\newblock \bibinfo{title}{Magnetic properties of alnico 5 and alnico 8 phases
  at the sequential stages of heat treatment in a field},
\newblock \bibinfo{journal}{Magnetics, IEEE Transactions on}
  \bibinfo{volume}{6} (\bibinfo{year}{1970}) \bibinfo{pages}{194--198}.
\bibitem[{Sergeyev and Bulygina(1969)}]{sergeyev.jap1969}
\bibinfo{author}{V.~Sergeyev}, \bibinfo{author}{T.~Bulygina},
\newblock \bibinfo{title}{Magnetic properties of alnico alloy phases and
  temperature instability of permanent magnets},
\newblock \bibinfo{journal}{Journal of Applied Physics} \bibinfo{volume}{40}
  (\bibinfo{year}{1969}) \bibinfo{pages}{1307--1307}.
\bibitem[{Iwama and Takeuchi(1974)}]{iwama.tjim1974}
\bibinfo{author}{Y.~Iwama}, \bibinfo{author}{M.~Takeuchi},
\newblock \bibinfo{title}{Spinodal decomposition in alnico 8 magnet alloy},
\newblock \bibinfo{journal}{Transactions of the Japan Institute of Metals}
  \bibinfo{volume}{15} (\bibinfo{year}{1974}) \bibinfo{pages}{371--377}.
\bibitem[{Iwama et~al.(1970)Iwama, Inagaki, and Miyamoto}]{iwama.tjim1970}
\bibinfo{author}{Y.~Iwama}, \bibinfo{author}{M.~Inagaki},
  \bibinfo{author}{T.~Miyamoto},
\newblock \bibinfo{title}{Effects of titanium in alnico 8-type magnet alloys},
\newblock \bibinfo{journal}{Transactions of the Japan Institute of Metals}
  \bibinfo{volume}{11} (\bibinfo{year}{1970}) \bibinfo{pages}{268--274}.
\bibitem[{Iwama(1967)}]{iwama.tjim1967}
\bibinfo{author}{Y.~Iwama},
\newblock \bibinfo{title}{Magnetic properties of alnico-type magnet alloys at
  elevated temperatures},
\newblock \bibinfo{journal}{Transactions of the Japan Institute of Metals}
  \bibinfo{volume}{8} (\bibinfo{year}{1967}) \bibinfo{pages}{18--25}.
\bibitem[{Tang et~al.(2015)Tang, Zhou, Kassen, Palasyuk, White, Dennis, Kramer,
  McCallum, and Anderson}]{tang.ieeetm2015}
\bibinfo{author}{W.~Tang}, \bibinfo{author}{L.~Zhou}, \bibinfo{author}{A.~G.
  Kassen}, \bibinfo{author}{A.~Palasyuk}, \bibinfo{author}{E.~M. White},
  \bibinfo{author}{K.~W. Dennis}, \bibinfo{author}{M.~J. Kramer},
  \bibinfo{author}{R.~W. McCallum}, \bibinfo{author}{I.~E. Anderson},
\newblock \bibinfo{title}{New alnico magnets fabricated from pre-alloyed
  gas-atomized powder through diverse consolidation techniques},
\newblock \bibinfo{journal}{IEEE Transactions on Magnetics}
  \bibinfo{volume}{51} (\bibinfo{year}{2015}) \bibinfo{pages}{1--3}.
\bibitem[{Thompson et~al.(2007)Thompson, Lawrence, Larson, Olson, Kelly, and
  Gorman}]{thompson.um2007}
\bibinfo{author}{K.~Thompson}, \bibinfo{author}{D.~Lawrence},
  \bibinfo{author}{D.~Larson}, \bibinfo{author}{J.~Olson},
  \bibinfo{author}{T.~Kelly}, \bibinfo{author}{B.~Gorman},
\newblock \bibinfo{title}{In situ site-specific specimen preparation for atom
  probe tomography},
\newblock \bibinfo{journal}{Ultramicroscopy} \bibinfo{volume}{107}
  (\bibinfo{year}{2007}) \bibinfo{pages}{131--139}.
\bibitem[{Miller and Russell(2007)}]{miller.sia2007}
\bibinfo{author}{M.~K. Miller}, \bibinfo{author}{K.~F. Russell},
\newblock \bibinfo{title}{Performance of a local electrode atom probe},
\newblock \bibinfo{journal}{Surface and Interface Analysis}
  \bibinfo{volume}{39} (\bibinfo{year}{2007}) \bibinfo{pages}{262--267}.
\bibitem[{Guo et~al.(2016)Guo, Garfinkel, Tucker, Haley, Young, and
  Poplawsky}]{guo.nanotech2016}
\bibinfo{author}{W.~Guo}, \bibinfo{author}{D.~A. Garfinkel},
  \bibinfo{author}{J.~D. Tucker}, \bibinfo{author}{D.~Haley},
  \bibinfo{author}{G.~A. Young}, \bibinfo{author}{J.~D. Poplawsky},
\newblock \bibinfo{title}{An atom probe perspective on phase separation and
  precipitation in duplex stainless steels},
\newblock \bibinfo{journal}{Nanotechnology} \bibinfo{volume}{27}
  (\bibinfo{year}{2016}) \bibinfo{pages}{254004}.
\bibitem[{Williams and Carter(1996)}]{williams.book1996}
\bibinfo{author}{D.~B. Williams}, \bibinfo{author}{C.~B. Carter},
  \bibinfo{title}{The Transmission Electron Microscope},
  \bibinfo{publisher}{Springer US}, \bibinfo{address}{Boston, MA},
  \bibinfo{year}{1996}, pp. \bibinfo{pages}{3--17}.
\bibitem[{Guo et~al.(2016)Guo, Sneed, Zhou, Tang, Kramer, Cullen, and
  Poplawsky}]{guo.mm2016}
\bibinfo{author}{W.~Guo}, \bibinfo{author}{B.~T. Sneed},
  \bibinfo{author}{L.~Zhou}, \bibinfo{author}{W.~Tang}, \bibinfo{author}{M.~J.
  Kramer}, \bibinfo{author}{D.~A. Cullen}, \bibinfo{author}{J.~D. Poplawsky},
\newblock \bibinfo{title}{Correlative energy-dispersive x-ray spectroscopic
  tomography and atom probe tomography of the phase separation in an alnico 8
  alloy},
\newblock \bibinfo{journal}{Microscopy and Microanalysis} \bibinfo{volume}{22}
  (\bibinfo{year}{2016}) \bibinfo{pages}{1251--260}.
\bibitem[{Laughlin and Soffa(1985)}]{laughlin.asmhandbook1985}
\bibinfo{author}{D.~E. Laughlin}, \bibinfo{author}{W.~Soffa},
\newblock \bibinfo{title}{Spinodal structures},
\newblock \bibinfo{journal}{ASM Handbook} \bibinfo{volume}{9}
  (\bibinfo{year}{1985}) \bibinfo{pages}{652--654}.
\bibitem[{Cahn(1961)}]{cahn.acta1961}
\bibinfo{author}{J.~W. Cahn},
\newblock \bibinfo{title}{On spinodal decomposition},
\newblock \bibinfo{journal}{Acta Metallurgica} \bibinfo{volume}{9}
  (\bibinfo{year}{1961}) \bibinfo{pages}{795--801}.
\bibitem[{Cahn(1963)}]{cahn.jap1963}
\bibinfo{author}{J.~W. Cahn},
\newblock \bibinfo{title}{Magnetic aging of spinodal alloys},
\newblock \bibinfo{journal}{Journal of Applied Physics} \bibinfo{volume}{34}
  (\bibinfo{year}{1963}) \bibinfo{pages}{3581--3586}.
\bibitem[{Zijlstra(1962)}]{zijlstra.zap1962}
\bibinfo{author}{H.~Zijlstra},
\newblock \bibinfo{title}{Magnetic annealing of “ticonal” g magnet steel},
\newblock \bibinfo{journal}{Z. Angew. Phys.} \bibinfo{volume}{14}
  (\bibinfo{year}{1962}) \bibinfo{pages}{251--253}.
\bibitem[{N\'eel(1947)}]{neel.cr1947}
\bibinfo{author}{N\'eel},
\newblock \bibinfo{journal}{Compt. rend.} \bibinfo{volume}{224}
  (\bibinfo{year}{1947}) \bibinfo{pages}{1488--1550}.

\end{thebibliography}

\end{document}